\documentclass[aps,superscriptaddress,amsfont,graphicx,nofootinbib,preprintnumbers]{revtex4-1}%
\UseRawInputEncoding
\usepackage{amsmath, amsthm, amssymb}
\usepackage{graphicx}
\usepackage{subfigure}
\usepackage{wrapfig}
\usepackage{color}
\usepackage{color,graphicx,epsfig}
\usepackage{ifpdf}
\usepackage{bm}
\usepackage[english]{babel}
\usepackage{braket}
\usepackage{hyperref}
\usepackage{enumerate}
\usepackage{url}
\usepackage{multirow}
\usepackage{tablefootnote} 
\usepackage{cases}

\usepackage{slashed}

\usepackage{changes}

\begin{document}
\title{Neutron Star Collapse From Accretion: a Probe of Massive Dark Matter Particles}

\author{Ning Liu}
\email{liuning@njnu.edu.cn}
\affiliation{Department of Physics and Institute of Theoretical Physics, Nanjing Normal University, Nanjing, 230021, China}

\author{
	Arvind Kumar Mishra 
	${\href{https://orcid.org/0000-0001-8158-6602}{\includegraphics[height=0.15in,width=0.15in]{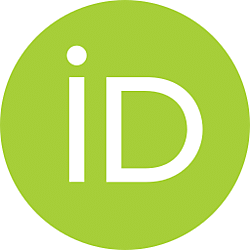}}}$ 
}
\email{arvindm215@gmail.com}
\affiliation{Department of Physics and Institute of Theoretical Physics, Nanjing Normal University, Nanjing, 230021, China}

\date{\today}

\begin{abstract}

We explore the multi-scatter capturing of the massive dark matter (DM) particle inside the neutron star via a momentum-dependent dark matter-nucleon scattering cross-section. We find that the capturing enhanced for the positive velocity and momentum transfer dependent DM-nucleon scattering in comparison with the constant cross-section case. Further, a large capture of the DM particles can be thermalized and lead to black hole formation and, therefore, destroy the neutron star. Using the observation of the old neutron star in the DM-dominated region, we obtain strong constraints on massive DM parameters.  
\end{abstract}
		

		\maketitle

\section{Introduction}
The various observations suggest that dark matter (DM) exists from its gravitational interaction with the baryons ~\cite{Bertone:2004pz}). The ongoing DM searches experiment assumes some form of the DM-baryon interactions; however, irrespective of current technological progress, the micro-physical properties of the DM particles remain an enigma \cite{Klasen:2015uma,Liu:2017drf,PerezdelosHeros:2020qyt}. 
The direct DM-detection experiments are sensitive for $\mathcal{O}(1-10^{4})$ GeV DM masses. However, if the DM particles are massive, i.e., $m_{\chi}> 10^{6}$ GeV, or lighter, $m_{\chi}<$ GeV, then during DM-nucleon collision, the energy transfer to the nucleus becomes small; therefore the detection experiment becomes insensitive to these masses.

In this work, we aim to explore the constraints on the massive dark matter, $m_{\chi}> 10^{6}$ GeV, from the neutron star observations. The neutron star provides a lucrative place to study the DM-nucleon interaction, thanks to their strong gravity and density, see a recent review \cite{Bramante:2023djs}. NS situated in the DM-dominated regions can be used to investigate the constraints in two ways: First, after capturing, DM particles become thermalized in the NS (via DM-nucleon scattering or DM-annihilation to the visible particles), which leads to heating of the NS and, therefore, observation of isolated of NS with high temperature can constraint the DM parameters \cite{Reisenegger:2006ky,Kouvaris:2007ay,Raj:2017wrv,Bell:2018pkk,Yanagi:2019vrr,Hamaguchi:2019oev,Bell:2020lmm,Garani:2020wge,Joglekar:2020liw,Fujiwara:2022uiq,Baryakhtar:2017dbj}. Second, after thermalization, the DM particles can form a self-gravitating core, which can collapse, form a black hole, and ultimately destroy the NS; therefore, observation of the old neutron star in a DM-dominated region (central region of galaxies) can used to probe the DM microphysics \cite{Goldman:1989nd, 1990PhLB..238..337G,Kouvaris:2010vv, deLavallaz:2010wp, McDermott:2011jp,Bramante:2014zca,Garani:2018kkd,Bramante:2013nma,Bramante:2013hn, Bell:2013xk, Guver:2012ba, Garani:2018kkd, Kouvaris:2018wnh,Lin:2020zmm,Garani:2021gvc,Singh:2022wvw}. However, most of the previous work has been done using a constant cross-section, without form factor, and the rest frame of the DM halo. Furthermore, incorporating the above effects into account, we obtained a strong constraint on the low-mass DM particles from the survival of NS \cite{Lu:2024kiz}.

The constraint on the massive DM has been explored from heating for both a single and multiscatter capture formalism \cite{Bell:2023ysh,Anzuini:2021lnv}. In previous works, \cite{McDermott:2011jp,Bramante:2014zca,Garani:2018kkd}, the single scatter capture formalism has been applied to put constraints on the massive DM particles from the survival of NS, specifically using constant DM-nucleon cross-section without form factor. However, for actual estimation, one needs to consider the multiscatter capture formalism for massive DM particles \cite{Bramante:2017xlb,Steigerwald:2022pjo,Leane:2023woh}.  For other methods to explore massive DM particles, please see Refs \cite{LHAASO:2022yxw,Song:2023xdk,Das:2024bed,LZ:2024psa,Bhattacharya:2024pmp}.

In this work, we explore the constraint on the massive DM particles from the neutron star survival. To explore the capture of massive DM particles on the neutron star, we adopt from the multiscatter capture approach. Here, we focus on the velocity-dependent and momentum-transfer dependent DM-nucleon cross-section, which, to our knowledge, was not explored for massive DM particles in the context of black hole formation (see, for low mass DM constraint in our previous work \cite{Lu:2024kiz}). In addition, we have also considered the general relativistic effect, the finite size effect of the nucleon, and the relative motion of the neutron star and galaxy frame in the capture rate.

Further, considering the power law parametrization of momentum and velocity dependent DM-nucleon cross-section, we found that the capture rate is enhanced with a positive power of velocity and momentum. We obtained that pulsar data provide the strongest constraint on DM parameter space (both for bosonic and fermionic DM case) for the model $\sigma\propto v^{2}_{\mathrm{rel}}$, where $ v_{\mathrm{rel}}$ is the relative velocity of DM and nucleon.
Using this model, pulsar observation strongly constrain bosonic DM mass range, $ 10^{6} ~\mathrm{GeV} \leq m_{\chi} \leq 6\times 10^{8}$ GeV, and fermionic DM masses, $ 10^{6} ~\mathrm{GeV} \leq m_{\chi} \leq  10^{10}$ GeV. We also report that although the DM accretion is higher for $\sigma\propto v^{4}_{\mathrm{rel}}$ model, a large thermalization time puts weak constraints on the DM parameters. Interestingly, a cross-section model with a large momentum dependency, i.e., $\sigma\propto q^{4}$ ($q$ represent the momentum transfer), provides no constraint on the DM parameters.

The arrangement of our work is as follows: In section \ref{sec:capture}, we present a brief discussion on the DM particle capture rate using a multiscatter capture process. In Section \ref{sec:BH}, we discuss the black hole formation from DM accretion. Further, we present DM-Nucleon cross-section and estimate the captured particles \ref{sec:accretion}. In Section \ref{sec:results}, we present and discuss our main results. Finally, in Section \ref{Sec:conc}, we conclude our work.
\section{Multiscatter Capture of DM particles on neutron star }
\label{sec:capture}
We assume that the neutron star is situated in a dark matter-rich environment and explore the capture of the massive DM particles to the neutron star. In the case of massive DM particles, an energy transfer via a single interaction between the DM particles and nucleons will be insufficient to capture the DM particles. Thus, a multiple scattering will be required \cite{Bramante:2017xlb,Steigerwald:2022pjo,Leane:2023woh}. Inside the NS, the number of times by which DM scatter with the nucleons can be obtained as the ratio of the typical diameter of the NS, $2R_{\mathrm{NS}}$ with the mean free path of the DM particles, $\lambda_{\mathrm{MFP}}$, given by  $N\sim 2R_{\mathrm{NS}}/\lambda_{\mathrm{MFP}}$, where $R_{\mathrm{NS}}$ is neutron star radius. Assuming $\lambda_{\mathrm{MFP}}\sim 1/(n_{\mathrm{N}}\sigma)$, one can obtain, $N\sim 2n_{\mathrm{N}}\sigma R_{\mathrm{NS}} $, where $\sigma$ is DM-nucleon scattering cross-section, and $n_{\mathrm{N}}$ is the nucleon density \cite{Bramante:2017xlb}.

Further, the probability of DM particles, $p_{\mathrm{N}}(\tau)$ participating in $N$ collisions with an optical depth, $\tau$  can be defined via Poisson distribution ($\mathrm{\tau,N}$) and given by \cite{Bramante:2017xlb}
\begin{equation}
 p_{\mathrm{N}}(\tau)=2\int^{1}_{0} y e^{-y\tau}  \frac{(y\tau)^{\mathrm{N}}}{\mathrm{N}!} dy~~,
 \label{eq:probability}
\end{equation}
where 
 $y$ is related to the incidence angle subtended by DM particles while entering the star.
The optical depth is defined as $\tau=\frac{3\sigma}{2\sigma_{\mathrm{sat}}}$, where $\sigma$ is the DM-nucleon cross-section and $\sigma_{\mathrm{sat}}$ is the saturation cross-section allowed during the scattering between the DM-nucleon interaction. In the context of NS, a saturation limit is dictated via geometrical cross-section, given by  $\sigma_{\mathrm{geo}}=\pi R^{2}_{\mathrm{NS}}/N_{\mathrm{B}}$ where, $R_{\mathrm{NS}}$ is the radius of NS, and $N_{\mathrm{B}}$ is total number of the baryons, respectively. For NS, considering, $R_{\mathrm{NS}}=10.6$ km and $N_{B}=1.7\times 10^{57}$, we obtain $\sigma_{\mathrm{sat}}=2\times 10^{-45}$ cm$^{2}$, which we assumed to remains a constant with DM mass    \footnote{Although we have assumed the saturation limit to be a constant quantity throughout the DM mass range, in general, it depends on the DM mass, see Refs. \cite{Baryakhtar:2017dbj,Bramante:2023djs,Ilie:2023lbi}. Therefore our estimation are somewhat conservative.}.

In the multiscatter capture process, the capture rate for $\mathrm{N}$ scatter, $C_{\mathrm{N}}$ is given by \cite{Bramante:2017xlb} \footnote{The derivation has been done  is the assumption of multi-capture with $v_{esc} (r)\simeq v_{esc} (R_{\mathrm{NS}}) $ and number density of the baryons are constant inside the neutron star, $n_{B}(r)=n_{B}$ see Ref. \cite{Bramante:2017xlb}}
\begin{equation}
C_{\mathrm{N}}=  \pi R^{2}_{\mathrm{NS}} p_{\mathrm{N}}(\tau)\int^{\infty}_{v_{\mathrm{esc}}} \frac{f(u)}{u^{2}}~ w^{3} g_{\mathrm{N}}(w)~dw~ .
\label{eq:Cnx}
\end{equation}
where $g_{\mathrm{N}}(w)$ corresponds to the probability of DM particles reducing their velocity lower than the escape velocity, $v_{\mathrm{esc}}$ after $N$ collisions, and $f(u)$ is the DM velocity distribution function.  In Eq. (\ref{eq:Cnx}), $u$ is the DM velocity at very large distance, and $ w=\sqrt{u^{2}+v^{2}_{\mathrm{esc}}}$. 
Therefore, the total capture rate after the $N$ collision is obtained as
\begin{equation}
C_{\mathrm{total}}=\sum^{\infty}_{\mathrm{N}=1} C_{\mathrm{N}}~.
\end{equation}
However, summation can be cut off at some value $\mathrm{N_{\max}}$ corresponding to condition, $p_{\mathrm{N_{\max}}}(\tau)\approx 0$. In our calculation, 
we approximate, $g_{\mathrm{N}}(w)=\Theta \left[v_{\mathrm{esc}}(1-\langle z_{i}\rangle \frac{\beta_{+}}{2})^{-N/2}-w\right]$, where, $\beta_{+}=\frac{4m_{\mathrm{N}} m_{\chi}}{(m_{\mathrm{N}}+m_{\chi})^{2}}$ in which, $m_{\chi}$ and $m_{N}$, represents the mass of the DM particles and nucleons, respectively \cite{Bramante:2017xlb}. Here we assume, $\langle z_{i}\rangle=1/2$, \cite{Bramante:2017xlb} however for more general estimation of $g_{\mathrm{N}}(w)$, see Ref. \cite{Dasgupta:2019juq}. 
Furthermore, the DM particle velocity distribution function is considered to be Maxwellian, given in the neutron star frame as \cite{1985ApJ...296..679P} 
\begin{equation}
 f_{0}(u) = \left(\frac{3}{2} \right)^{\frac{3}{2}} \frac{4}{\sqrt{\pi}} \frac{\rho_{\chi}}{m_{\chi}} \frac{u^{2}}{u^{3}_{0}}e^{-\frac{3}{2}\frac{u^{2}}{u^{2}_{0}}}~,
 \label{eq:dist}
 \end{equation}
 where $\rho_{\chi}$, and $u_{0}$ represent energy density, mass, and average velocity of the DM particles, respectively.

Furthermore, as on the surface of the neutron star, the DM velocity is high, so during the interaction, DM imparts large energy to the nucleons, see Ref.\cite{Bell:2020obw}. Therefore, in this analysis, we consider the finite-size effect of the nucleon during the DM-Nucleon interaction. Here, we adopt a conservative approach and assume the dipole form factor for DM-nucleon interaction (considered in the next subsection), given by \cite{Ema:2020ulo,Lu:2024kiz}
\begin{equation}
F(q^{2})=\frac{\Lambda^{4}}{(q^{2}+\Lambda^{2})^{2}} 
\label{eq:FF}
\end{equation}
where $q\simeq \sqrt{2}\left(\frac{m_{\chi}m_{\mathrm{N}}}{m_{\chi}+m_{\mathrm{N}}}\right)v_{\mathrm{esc}}$ is momentum transfer during the DM-nucleon collision and $m_{\mathrm{N}}$ is the nucleon mass. Further, we assume $\Lambda\simeq 0.25$ GeV throughout our analysis \cite{Lu:2024kiz}. Furthermore, in our computation, we use the general relativistic (GR) correction, which modifies the capture rate and escapes velocity. So, using the GR correction, the capture rate and escape velocity is modifies via \cite{Goldman:1989nd,Bramante:2017xlb}
\begin{equation}
 C_{\mathrm{N}}\rightarrow \frac{C_{\mathrm{N}}}{(1-v^{2}_{\mathrm{esc})}} \quad \mathrm{and} \quad  v_{\mathrm{esc}}\rightarrow \sqrt{2\left[ 1-(1-v^{2}_{\mathrm{esc}})^{\frac{1}{2}}\right]}.
\end{equation}

\section{Black hole formation from DM accretion}
\label{sec:BH}
After the accretion, the DM particles start interacting with the neutron star material and may get thermalized. In the context of the multi-capture process, the interaction of the DM particles with the neutron star is complex because, at each step, the DM particle transfers less momentum, and hence, the estimation of the thermalization time is complex.  Now, we consider the simplified approach and assume the form of the thermalization time scale as \cite{Garani:2018kkd}
\begin{equation}
t_{\mathrm{th}}  \simeq   10700\frac{\mu}{(1+\mu)^{2}}\left(\frac{10^{5} \mathrm{K}}{T_{\mathrm{NS}}}\right)^{2}\left(\frac{10^{-45} \mathrm{cm}^{2}}{\sigma}\right) \quad \mathrm{years}.
\label{eq:th}
\end{equation}
where, $\mu=m_{\chi}/m_{\mathrm{N}}$, and $\sigma$ is DM-nucleon scattering cross-section. When the thermalization time scale is smaller than the lifetime of the NS ($t_{\mathrm{th}}<t_{\mathrm{NS}}$), DM particles will be thermalized inside the NS and lie inside the  thermal radius given by 
\begin{equation}
r_{\mathrm{th}}=
\left( \frac{9T_{\mathrm{NS}}}{8\pi G_{\mathrm{N}} \rho_{\mathrm{B}} m_{\chi}}\right)^{\frac{1}{2}} ~~.
\end{equation}
where $G_{\mathrm{N}}$ is the Newton's gravitational constant,  and $\rho_{\mathrm{B}}$ is the baryon energy density. For a given density and temperature of the NS, $r_{\mathrm{th}}\propto
\sqrt{\frac{1}{ m_{\chi}}}$, which implies that the thermal radius is smaller for massive DM particles and vice versa. 
Further, DM particles start self-gravitating whenever
\begin{equation}
\rho_{\chi}>\rho_{\mathrm{B}} \quad \mathrm{for} \quad  r \leq r_{\mathrm{th}}.
\end{equation}
Therefore, with the help of the above condition, we define the DM particles dictated via $N_{\mathrm{self}}$, necessary for the self-gravitation of the DM particles inside the NS as
\begin{equation}
N_{\mathrm{self}}\simeq  4.8\times 10^{41} \left( \frac{1.4\times 10^{14} \mathrm{gm}/\mathrm{cm}^{3}}{\rho_{B}}\right)^{\frac{1}{2}}   \left( \frac{100 \mathrm{GeV}}{m_{\chi}}\right)^{\frac{5}{2}}  \left( \frac{T_{\mathrm{NS}}}{10^{5} \mathrm{K}}\right)^{\frac{3}{2}}~~.
\label{eq:Nself}
\end{equation}
So the condition for DM self-gravitation suggest, $N_{\chi}\geq N_{\mathrm{Self}}$.

Furthermore, when the number of DM particles becomes higher than those defined by the Chandrashekhar limit, then self-gravitating DM particles collapse and form a black hole. In the case of fermionic dark matter (FDM) particles, gravitational collapse requires gravity to surpass the fermi pressure. For non-interacting fermionic DM particles, the Chandrashekhar  limit is obtained as \cite{McDermott:2011jp}
\begin{equation}
N^{\mathrm{Fermions}}_{\mathrm{Chandra}}\simeq 1.8\times 10^{51} \left( \frac{100 \mathrm{GeV}}{m_{\chi}}\right)^{3} ~.
\label{eq:Fchand}
\end{equation}
From the above equation, it is important to emphasize that the massive DM particles form low-mass BH and vice versa.
In this case, the BH formation occurs when DM particles start self-gravitating, i.e., $N_{\chi}\geq N_{\mathrm{Self}}$, and captured DM particles are higher than the Chandrashekhar limit,i.e.,$N_{\chi}>N^{\mathrm{Fermions}}_{\mathrm{Chandra}}$. The mass of formed BH is obtained as $M_{\mathrm{BH}}\sim m_{\chi}N^{\mathrm{Fermions}}_{\mathrm{Chandra}}$, and given by
\begin{equation}
M_{\mathrm{BH}}\simeq    1.8\times 10^{53}\left( \frac{100 \mathrm{GeV}}{m_{\chi}}\right)^{2}\mathrm{GeV}.
\label{eq:BHmass}
\end{equation}
From the above equation, it is important to emphasize that the massive DM particles form low-mass BH and vice versa. 
Furthermore, to form a BH in the bosonic dark matter (BDM) particle case, gravity must overcome pressure originating from the uncertainty relation. For non-interacting bosons, the Chandrashekhar limit is given by \cite{McDermott:2011jp}
\begin{equation}
N^{\mathrm{Bosons}}_{\mathrm{Chandra}}\simeq 1.5\times 10^{34} \left( \frac{100 \mathrm{GeV}}{m_{\chi}}\right)^{2} ~.
\label{eq:Bchand}
\end{equation} 
In bosonic DM case, BH forms as soon as DM starts self-gravitating, 
\begin{equation}
N_{\chi}\geq N_{\mathrm{Self}}, ~  \mathrm{for} ~ m_{\chi}\leq 10^{21} \left( \frac{\rho_{B}}{1.4\times 10^{14} \mathrm{gm}/\mathrm{cm}^{3}}\right)  \left( \frac{10^{5}\mathrm{K}}{T_{\mathrm{NS}}}\right)^{3} ~\mathrm{GeV}~~.
\end{equation}
In this case, the mass of formed BH is obtained as $M_{\mathrm{BH}}\sim m_{\chi}N_{\mathrm{Self}}$, and explicitly given by
\begin{equation}
M_{\mathrm{BH}}\simeq 4.8\times 10^{43}\left( \frac{1.4\times 10^{14} \mathrm{gm}/\mathrm{cm}^{3}}{\rho_{B}}\right)^{\frac{1}{2}} \left( \frac{100 \mathrm{GeV}}{m_{\chi}}\right)^{\frac{1}{2}}  \left( \frac{T_{\mathrm{NS}}}{10^{5} \mathrm{K}}\right)^{\frac{3}{2}}~~\mathrm{GeV}.
\label{eq:BHmass}
\end{equation}
From the above equation, we find that the massive DM particles and low neutron star temperature form low-mass BH and vice versa.

After its formation, BH starts growing via accreting the surrounding material and also reducing its mass via Hawking evaporation. The BH mass evolves  as \cite{McDermott:2011jp}
\begin{equation}
 \frac{dM_{\mathrm{BH}}}{dt}\simeq 4\pi \lambda_{s} \left( \frac{G_{\mathrm{N}}M_{\mathrm{BH}}}{v^{2}_{s}}\right)^{2}\rho_{\mathrm{B}} v_{s}+\left(\frac{dM_{\mathrm{BH}}}{dt}\right)_{\mathrm{DM}} -\frac{1}{1560\pi G^{2}_{\mathrm{N}} M^{2}_{\mathrm{BH}}}~~,
\label{eq:hawking}   
\end{equation}
where $\lambda_{s}$ and $v_{s}$ are the accretion eigenvalue and sound speed. In the r.h.s, the first and second term corresponds to the accretion rate obtained from baryon (Bondi-Hoyle accretion rate) and DM particles, and the last term represents Hawking evaporation. Assuming the DM accretion contribution is small, Eq.~(\ref{eq:hawking}) suggests an initial critical BH mass given as $M^{\mathrm{crit}}_{\mathrm{BH}}=1.2\times 10^{37}$ GeV. Whenever formed BH with initial mass,$M^{\mathrm{ini}}_{\mathrm{BH}}$ is larger than the critical mass, i.e., $M^{\mathrm{ini}}_{\mathrm{BH}}>M^{\mathrm{crit}}_{\mathrm{BH}}$, BH will grow via accretion and destroy the host NS. However, in the opposite case, i.e., $M^{\mathrm{ini}}_{\mathrm{BH}}<M^{\mathrm{crit}}_{\mathrm{BH}}$, black hole will evaporate and decrease its mass with the time. 

Furthermore, the value of the critical mass of the formed BH can be applied to estimate an upper limit on the mass of DM particles, which can be probed from the BH formation in the core of NS. The initial formed BH mass formed by fermionic DM particles,$M_{\mathrm{BH}}\sim m_{\chi}N^{\mathrm{Fermions}}_{\mathrm{Chandra}}$, and bosonic DM particles via $M_{\mathrm{BH}}\sim m_{\chi}N_{\mathrm{Self}}$. Therefore assuming that BH grows via accreting the NS materials, we obtain the $m_{\chi}<2.5\times 10^{8}$ GeV (for $T_{\mathrm{NS}}=10^{7}$ Kelvin) for BDM particles, and $m_{\chi}<10^{10}$ GeV for the FDM particles. We emphasize that these constraints are obtained using the assumption that formed BH is growing via accreting the NS material and not considered DM accretion. Therefore, when DM accretion is taken into account (or NS temperature is changed for the BDM case), the constraint will be modified.

 \section{DM Accretion from non-constant DM-Neucleon elastic scattering}
 \label{sec:accretion}
In this work, we explore the DM-Nucleon interaction, which is dependent on relative velocity and momentum transfer during the collision \cite{Chang:2009yt,Fan:2010gt,Kumar:2013iva,Vincent:2015gqa,Vincent:2016dcp,Ooba:2019erm,Maamari:2020aqz,Buen-Abad:2021mvc,Boddy:2022tyt}, which are investigated in contest of the DM capturing on sun~\cite{Guo:2013ypa,Vincent:2013lua,Vincent:2014jia,Vincent:2015gqa,Vincent:2016dcp,Busoni:2017mhe}, white dwarfs~\cite{Steigerwald:2022pjo} and neutron stars~\cite{Bell:2018pkk,Bell:2020lmm,Garani:2020wge,Joglekar:2020liw,Fujiwara:2022uiq}. 
 In this work, we parameterize the relative velocity-dependent cross-section (VDCS) as a power law form, given by \cite{Vincent:2015gqa,Lu:2024kiz}
\begin{equation}
\sigma^{\mathrm{vd}}_{\chi\mathrm{N}}(v_{\mathrm{rel}})=\sigma_{\chi \mathrm{N}}\left(\frac{v_{\mathrm{rel}}}{u_{0}}\right)^{2\alpha}\simeq\sigma_{\chi \mathrm{N}}\left(\frac{w}{u_{0}}\right)^{2\alpha},
\label{eq:vdcs}
\end{equation}
where, $v_{\mathrm{rel}}$ is the DM-nucleon relative velocity, and $u_{0}$ represent the normalized velocity. Here $\alpha$ depends on the DM models, and we specifically focus on the positive velocity dependence cases, i.e., $ \alpha=1, \mathrm{and}~2$.  This kind of interaction is called p-wave ($\alpha=1$), and d-wave ($\alpha=2$) can be originated for the cases when initial state particles have relative angular momentum of 1 or 2 units, see Refs. \cite{Vincent:2015gqa,Kumar:2013iva},\cite{Sigurdson:2004zp,Dvorkin:2013cea,Boddy:2022tyt}.  Here we consider the typical value, $u_{0}=220$ km$/$sec \cite{Vincent:2015gqa,Lu:2024kiz}.

Further, we parameterize the momentum-dependent cross-section (MDCS) in the power law form as\cite{Vincent:2015gqa,Lu:2024kiz}
\begin{equation}
\sigma^{\mathrm{md}}_{\chi\mathrm{N}}(q)=\sigma_{\chi \mathrm{N}}\left(\frac{q}{q_{0}}\right)^{2\beta},
\label{eq:mdcs} 
\end{equation}
where $q$ is the momentum transfer and  $q_{0}$ is a normalized momentum. The momentum dependent cross-section can originate from various mechanisms such as dipole interaction between the DM and baryons, parity violating DM-baryon interaction, see Refs. \cite{Feldstein:2009tr,Chang:2009yt,Chang:2010en,Fan:2010gt,Kumar:2013iva,Vincent:2015gqa}. Further, for calculation purposes, we assume, $q_{0}=40$ MeV \cite{Vincent:2015gqa,Lu:2024kiz}).

Having been equipped with the form of the DM-baryon interaction, we will now calculate the probability and DM capture rate using the multicapture process.  This can be obtained by using Eq.(\ref{eq:probability}) and Eq.(\ref{eq:Cnx}). We have estimated the multiplescatter capture rate using constant, MDCS and VDCS in Appendices \ref{constant},\ref{momdep}, and \ref{veldep}, respectively. Below, we will briefly summarize the  results:

\textbf{Capture rate in constant cross-section case:} For constant cross-section case, the expression of probability is obtained after using Eq.(\ref{eq:probability}) as
\begin{equation}
P^{\mathrm{cont}}_{\mathrm{N}}(\tau_{0})=\frac{2}{\tau^{2}_{0} N!} \bigg[\Gamma(2+N)-\Gamma(2+N,\tau_{0})\bigg]
\label{eq:pconst}
\end{equation}
where, $\tau_{0}=\frac{3\sigma_{\chi \mathrm{N}} F(q^{2})}{2\sigma_{\mathrm{sat}}}$, and $\Gamma(a,b)$ represent the incomplete Gamma function. Therefore using the above probability, $C_{\mathrm{N}}$ can be obtained as
 \begin{equation}
  C^{\mathrm{cont}}_{\mathrm{N}}(\tau_{0})=   \frac{R^{2}}{1-v^{2}_{\mathrm{esc}}}\frac{n_{\chi}}{6u_{0}}P^{\mathrm{cont}}_{\mathrm{N}}(\tau_{0})f(v_{\mathrm{N}},\eta)\Theta(v_{\mathrm{N}}-v_{\mathrm{esc}})
   \label{eq:cnconst}
 \end{equation}
where, $v_{\mathrm{N}}=v_{\mathrm{esc}}\left(1-\frac{\beta_{+}}{2} \right)^{-\frac{N}{2}}$ is the DM velocity after $N$ scattering \cite{Bramante:2017xlb}. Further, the total capture rate after $N$ collision is given by $C^{\mathrm{cont}}_{\mathrm{tot}}(\tau_{0})=\sum^{\infty}_{\mathrm{N}=1}C^{\mathrm{cont}}_{\mathrm{N}}(\tau_{0})$. In this case, an analytic expression of the total capture rate is obtained using various analytic approximations, see Refs. \cite{Ilie:2020vec,Ilie:2023lbi,Ilie:2024sos}. Further, to calculate the total collected DM particles, we considered DM particles to be asymmetric (neglecting DM annihilation). In this assumption, the number of collected DM particles by neutron star throughout its lifetime is given by $N_{\chi}(t_{\mathrm{NS}})=C^{\mathrm{cont}}_{\mathrm{tot}}(\tau_{0}) t_{\mathrm{NS}}$. Here, we will not apply the analytical approximations and calculate $N_{\chi}(t_{\mathrm{NS}})$ numerically.

\textbf{Capture rate in MDCS case:} For MDCS, the optical depth is given by $\tau_{\mathrm{md}}=\tau_{0}\left(q/q_{0}\right)^{2\beta}$. So, the probability after $N$ collision is obtained as
\begin{equation}
 P^{\mathrm{MDCS}}_{\mathrm{N}}(\tau_{0},q)=\frac{2}{\tau^{2}_{0} N!} \left( \frac{q}{q_{0}}\right)^{-4\beta}
 \bigg[\Gamma(2+N)-\Gamma\left(2+N,\tau_{0} ( q/q_{0})^{2\beta}\right)\bigg]~. 
 \label{eq:pmdcs}
 \end{equation}
The form of $C_{\mathrm{N}}$ in the MDCS case is given by 
\begin{equation}
  C^{\mathrm{MDCS}}_{\mathrm{N}}(\tau_{0},q)= \frac{R^{2}}{1-v^{2}_{\mathrm{esc}}}\frac{n_{\chi}}{6u_{0}}P^{\mathrm{MDCS}}_{\mathrm{N}}(\tau_{0},q)f(v_{\mathrm{N}},\eta)\Theta(v_{\mathrm{N}}-v_{\mathrm{esc}})
   \label{eq:cnmdcs}
 \end{equation}
  Further, the total capture rate after $N$ collision is given by $C^{\mathrm{MDCS}}_{\mathrm{tot}}(\tau_{0},q)=\sum^{\infty}_{\mathrm{N}=1}C^{\mathrm{MDCS}}_{\mathrm{N}}(\tau_{0},q)$. Further, assuming the DM particles to be asymmetric, the total captured DM particles in the MDCS case is given by $N_{\chi}(t_{\mathrm{NS}})=C^{\mathrm{MDCS}}_{\mathrm{tot}}(\tau_{0},q) t_{\mathrm{NS}}$. In this case, we will calculate $N_{\chi}(t_{\mathrm{NS}})$ numerically.  

\textbf{Capture rate in VDCS case:} For VDCS, the optical depth is given by $\tau_{\mathrm{vd}}=\tau_{0}\left(w/u_{0}\right)^{2\alpha}$. So, the probability after $N$ collision is obtained as
\begin{equation}
 P^{\mathrm{VDCS}}_{\mathrm{N}}(\tau_{0},w)=\frac{2}{\tau^{2}_{0} N!} \left( \frac{w}{u_{0}}\right)^{-4\alpha}
 \bigg[\Gamma(2+N)-\Gamma\left(2+N,\tau_{0} ( w/u_{0})^{2\alpha}\right)\bigg]~. 
 \label{eq:pvdcs}
 \end{equation}
Using the above probability after simplification (see appendix \ref{veldep}), we obtain an analytic expression for the capture rate as
  \begin{equation}
  C^{\mathrm{VDCS}}_{\mathrm{N}}(\tau_{0},w_{0})= \frac{R^{2}}{1-v^{2}_{\mathrm{esc}}}\frac{n_{\chi}}{6u_{0}}P^{\mathrm{VDCS}}_{\mathrm{N}}(\tau_{0},w_{0})f(v_{\mathrm{N}},\eta)\Theta(v_{\mathrm{N}}-v_{\mathrm{esc}})
   \label{eq:cnvdcs}
 \end{equation}
In this case, the total capture rate after $N$ collision is obtained as $C^{\mathrm{VDCS}}_{\mathrm{tot}}(\tau_{0},w_{0})=\sum^{\infty}_{\mathrm{N}=1}C^{\mathrm{VDCS}}_{\mathrm{N}}(\tau_{0},w_{0})$. Further, in the assumption that the DM particles are asymmetric, the total captured DM particles in this case is given by $N_{\chi}(t_{\mathrm{NS}})=C^{\mathrm{VDCS}}_{\mathrm{tot}}(\tau_{0},w_{0}) t_{\mathrm{NS}}$. In this case, we will calculate $N_{\chi}(t_{\mathrm{NS}})$ numerically. 

\begin{figure*}
 \includegraphics[height=2.5in,width=3.2in]{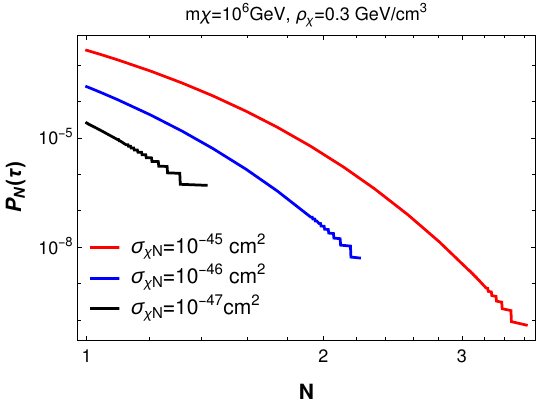}\hfil
 \includegraphics[height=2.5in,width=3.2in]{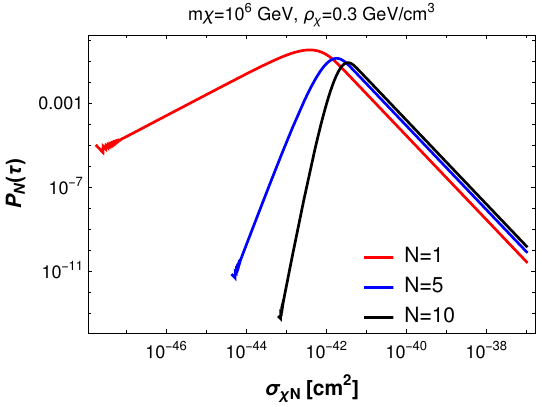}\par
\caption{The capturing probability as a function of the number of the DM-Nucleon scattering, $N$  (Left Panel), and cross-section (Right panel) for constant DM-Nucleon scattering cross-section.   }
	\label{fig:pnconst}%
\end{figure*}
\begin{figure*}
   \includegraphics[height=2.5in,width=3.2in]{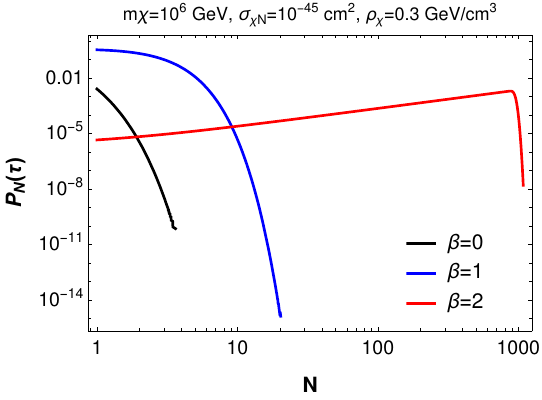}\hfil
 \includegraphics[height=2.5in,width=3.2in]{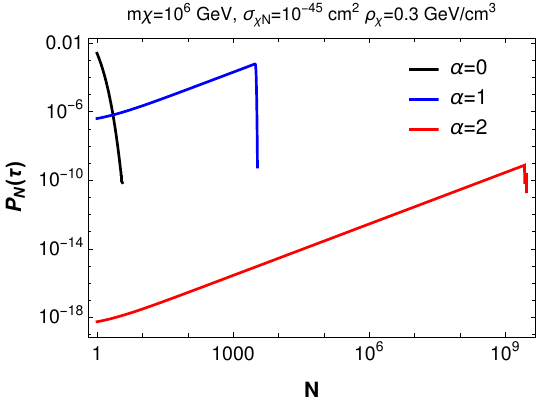}\par
\caption{The probability of multicapture in the momentum-dependent (Left panel) and velocity-dependent (Right panel) DM-Nucleon scattering cross-section.   }
\label{fig:pmdvd}%
\end{figure*}
Furthermore, the estimation of total accreted DM particles using the general expression of capture rate and probability is tedious. Therefore, we try to restrict our calculation up to a certain accuracy as follows:  We calculate the total sum up to $N=N_{\mathrm{max}}$, where $N_{\mathrm{max}}\leq \tau$ \cite{Ilie:2020vec}. Further, we use a simplified expression of probability in some extreme cases of optical depth. For a single scatter ($\tau\ll 1$) and multiscatter ($\tau\gg 1$), the probability is approximated as  \cite{Bramante:2023djs}
\begin{subnumcases}
{P_{\mathrm{N}} (\tau)\approx} 
\frac{2\tau^{\mathrm{N}}}{N!(N+2)}+\mathrm{O}(\tau^{\mathrm{N}+1}), \quad if~~ \tau\ll 1\label{eq:CtotSSHM}\\
\frac{2}{\tau^{2}}\left(N+1\right) \Theta \left(\tau-N\right), \quad if~~ \tau\gg 1 \label{eq:CtotSSLM}
\end{subnumcases}
Therefore, using the capture rate expression estimated above, we calculate the accreted DM particles for constant, MDCS, and VDCS cases numerically.

\section{Results and discussions}
\label{sec:results}
In this Section, we will present our results. For calculation purposes, we assume the NS parameters as $M_{\mathrm{NS}}=1.44 M_{\odot}$, $R_{\mathrm{NS}}=10.6$ km, $N_{\mathrm{B}}=1.7\times 10^{57}$, and $v_{\mathrm{esc}}=1.8\times 10^{5}$ km$/$sec \cite{McDermott:2011jp}, otherwise specified explicitly. 
\subsection{Effect of DM-Nucleon interactions on the multiscatter probability}
In Fig.~\ref{fig:pnconst}, we plot the probability of the DM multiscatter capture as a function of a number of scattering, $N$, (left panel) and cross-section (right panel). It can be seen that the probability decreases as the number of scattering $N$ increases and becomes zero for some value of $N$, say $N_{\mathrm{max}}$, i.e., $p_{N_{\mathrm{max}}}(\tau)\approx 0$. Further, fixing the DM parameters, $m_{\chi}=10^{6}$ GeV,$\rho_{\chi}=0.3$GeV$/$cm$^{3}$ and $N_{\mathrm{max}}$ takes values $1$ and $2$, and $3$ corresponding to $\sigma_{\chi\mathrm{N}}=10^{-47}$cm$^{2}$, $\sigma_{\chi\mathrm{N}}=10^{-46}$cm$^{2}$, and $\sigma_{\chi\mathrm{N}}=10^{-45}$cm$^{2}$, respectively. 
 Also, increments in the cross-section lead to an enhanced probability and, therefore, enlarge $N_{\mathrm{max}}$. 
Further, the right panel of Fig~\ref{fig:pnconst} suggests that
depending on the $N$, the probability becomes max at some $\sigma_{\mathrm{crit}}$ and vanishes on low DM-Nucleon scattering cross-section, $\sigma_{\mathrm{min}}$ such that $\sigma_{\mathrm{min}}<\sigma_{\mathrm{crit}}$ . Nevertheless, increasing the number of scattering will shift the peak probability position and $\sigma_{\mathrm{min}}$ on a comparably higher cross-section.  Furthermore, we also find that increasing the DM masses decreases the probability.

\begin{figure*}
 \includegraphics[height=2.5in,width=3.2in]{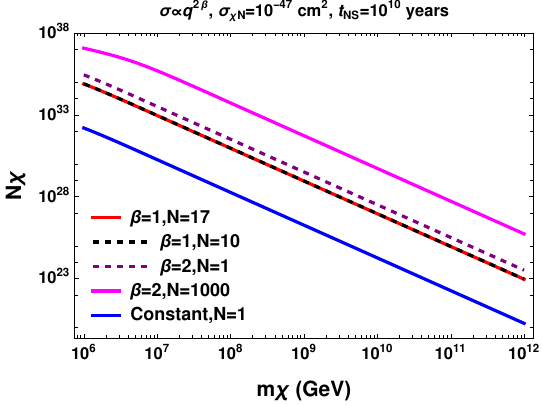}\hfill
  \includegraphics[height=2.5in,width=3.2in]{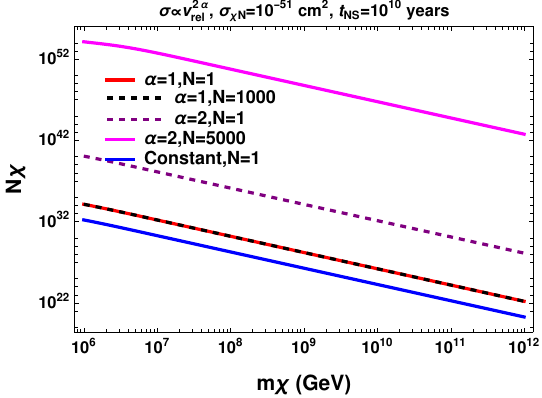}\par
\caption{The accreted number of the DM particles via multiscatter process. The Left and Right panels correspond to the MDCS and VDCS cases, respectively.    }
	\label{fig:captrate}%
\end{figure*}

In Fig.~\ref{fig:pmdvd}, we plot the probability of the DM capture using MDCS, $\sigma\propto q^{2\beta}$ (Left panel) and VDCS, $\sigma\propto v^{2\alpha}_{\mathrm{rel}}$ (Right panel) cases.  From the Left panel of Fig.~\ref{fig:pmdvd}, it can be seen that the probability decreases as the number of scattering $N$ increases and becomes zero for some value $N_{\mathrm{max}}$ as discussed previously. However, $\beta=2$ shows an interesting feature that probability increases first and decreases sharply after a large scattering. For our parametrization and fixing the other DM parameters, $m_{\chi},\rho_{\chi}$ and $\sigma_{\chi\mathrm{N}}$, $N_{\mathrm{max}}$ takes values $6,30$ and $1000$ corresponding to $\beta= 0,1,$ and $2$, respectively.   Further, in the VDCS case (Left panel of Fig.~\ref{fig:pmdvd}), DM capturing inside the NS requires more scattering, $N_{\mathrm{max}}$. For example, keeping DM parameters fixed, $m_{\chi}=10^{6}$ GeV, $\rho_{\chi}=0.3\mathrm{GeV}/$cm$^{3}$ and $\sigma_{\chi\mathrm{N}}=10^{-45}$cm$^{2}$, DM capturing in $\alpha=2$ case required $\sim 10^{9}$ scattering. Further, for constant cross-section, the probability decreases promptly, leading to a smaller $N_{\mathrm{max}}$ value in comparison with the MDCS and VDCS cases. It is important to emphasize that the value of $N_{\mathrm{max}}$ depends on the DM parameters; therefore, changing it will modify the $N_{\mathrm{max}}$ value.

\begin{figure*}
   \includegraphics[height=3in,width=4in]{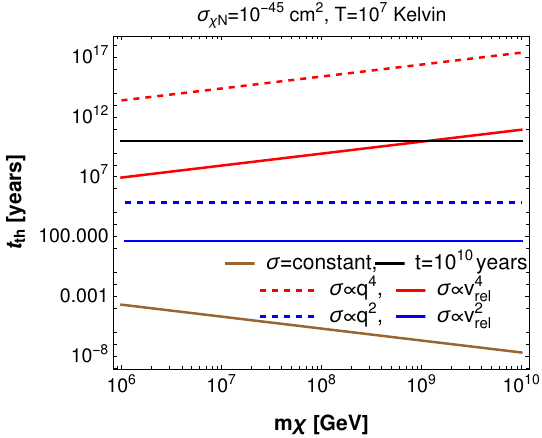}\hfil
\caption{The thermalization time scale as a function of DM mass.  }
\label{fig:temcapt}%
\end{figure*}
\subsection{Accreted DM particles via multi-capture process}
 Fig.~\ref{fig:captrate} shows the accreted DM particles via multi-scatter capture for MDCS (Left panel), and VDCS (Right panel) cases. We found that for both cases, the number of the accreted DM particles, $N_{\chi}$, enhanced with the DM scattering, but decreased with the DM mass. For the MDCS case, $N_{\chi}$ increases with the positive power of cross-section, $\beta>0$ and number of scattering, $N$. Here we see that for $\beta=1$, $N_{\chi}$ is same for $N=1$ and $N=10$. This is because for such a small DM-neutron scattering and also parameters used in the calculation (see Left panel of Fig.~\ref{fig:captrate}), DM capturing happens through single scattering, and therefore  $N_{\chi}$ remains constant for $N>1$. For the $\beta=2$ case, DM capturing becomes possible via multiscattering, and therefore, $N_{\chi}$ increases with $N$ until it gets saturated at $N_{\mathrm{max}}=1000$.  Furthermore, in the VDCS case (Right panel of Fig.~\ref{fig:captrate}) and for $\alpha=1$, the number of captured DM particles is equal for $N=1$ and $N=1000$ (same reason as previously). For the strong VDCS case, $\alpha=2$, DM particles get captured via multiscattering, and therefore, $N_{\chi}$ increases till its saturation at $N_{\mathrm{max}}=5000$. From  Fig.~\ref{fig:captrate}, we also point out that for positive MDCS and VDCS cases, the number of captured DM particles is higher than the constant cross-section case.

\begin{figure*}
  \includegraphics[height=3in,width=3.2in]{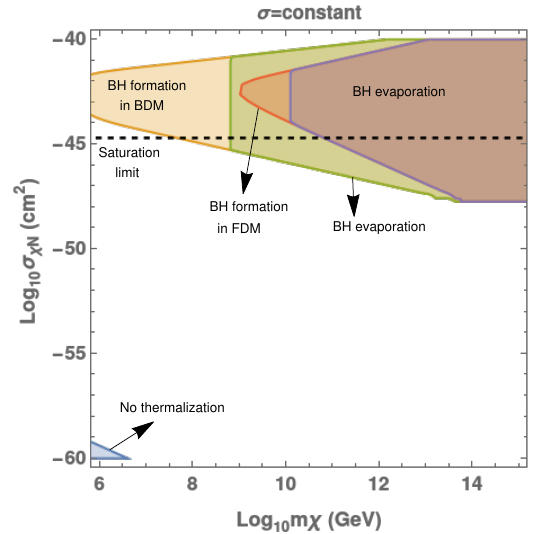}\par
\caption{Constraints on the DM-nucleon scattering cross-section for constant cross-section obtained using the pulsar data J2124-3858. Here, orange, red, and sky-blue color regions are excluded from the BH formation via bosonic DM, fermionic DM, and non-thermalization of the captured DM particles. Further, olive and brown color regions represent the BH evaporation formed from BDM and FDM, respectively. The black dashed line is the saturation limit of the DM-Nucleon scattering cross-section.} 
\label{fig:constconst}%
\end{figure*}

\subsection{Thermalization time scale}
To estimate a DM thermalization time for non-constant DM-nucleon interaction, we assume as follows: In VDCS and MDCS case, we replace $\sigma$ via Eq.(\ref{eq:vdcs}), and Eq.(\ref{eq:mdcs}), and the relative velocity is assumed to be the thermal velocity inside the NS, i.e., $v_{\mathrm{rel}}\simeq v_{\mathrm{th}}=\sqrt{3T_{\mathrm{NS}}/m_{\chi}}$. With this simplification, we plot the thermalization time scale for our various parameterizations in Fig.~\ref{fig:temcapt}. We find that for constant and $\sigma\propto v^{2}_{\mathrm{rel}}$, $\sigma\propto q^{2}$, $t_{\mathrm{th}}$ is remains constant but increases linearly with mass, i.e., $t_{\mathrm{th}}\propto m_{\chi}$. This behavior is evident from Eq.(\ref{eq:vdcs}), and Eq.(\ref{eq:mdcs}). It can be seen that for a few cases, i.e. 
$\sigma\propto v^{2}_{\mathrm{rel}},\sigma\propto v^{4}_{\mathrm{rel}}$ and $\sigma\propto q^{2}$, DM quickly thermalized inside the NS, and therefore, these cases are interesting in the context of the BH formation. However, for the $\sigma\propto ~q^{4}$ case, the thermalization scale is quite large; therefore, DM does not thermalize in the typical lifetime of the NS.

\subsection{Constraining massive DM from NS collapse}
\label{subsec:vdconst}

Now, we will constrain the DM-nucleon parameter space represented via $\sigma_{\chi \mathrm{N}} -m_{\chi} $ using the survival of the neutron star situated in the DM-dominated region. We consider the cold and old neutron stars whose distance from the galaxy center, central temperature, and age are estimated accurately. For this purpose, we consider pulsar J2124-3858 situated at a distance $270$ pc with surface temperature, $T_{e}<4.6\times 10^{5}$K (central temperature, $T_{c}=2.5\times 10^{7}$K \footnote{The central temperature can be estimated using the surface data from the analytic formula discussed in Ref.~\cite{1982ApJ...259L..19G} .}), $t_{\mathrm{age}}\sim 7.81\times10^{9} $ years. As this pulsar is close to the sun, we may approximate $u_{\mathrm{s}}\simeq u_{{\odot}}= 230$ km$/$sec \cite{Guo:2013ypa}, and $\rho_{\chi}=0.3\mathrm{GeV~ cm^{-3}}$, where $u_{{\odot}}$ represent the sun's speed relative to galactic halo.

Fig~\ref{fig:constconst} shows a constraint on the DM-nucleon scattering cross-section for a constant cross-section using the pulsar data J2124-3858. Here, orange and red colors are excluded from the BH formation via bosonic and fermionic DM particles. The olive and brown color regions represent the BH evaporation regions for bosonic and fermionic DM particles, and in these regions, the constraint is relaxed. Further, sky-blue regions are excluded via non-thermalization of the captured DM particles, i.e., $t_{\mathrm{th}}>t_{\mathrm{NS}}$. The black dashed line is the saturation limit of the DM-Nucleon scattering cross-section, and above this line, the constraints are not valid.

From Fig~\ref{fig:constconst}, we see that for BDM, the number of the accreted DM particles is high enough to form the BH across all the DM mass ranges. The higher DM particles form low mass BH, so for sufficiently large DM masses, formed BH masses become smaller than the critical mass, i.e., $M^{\mathrm{ini}}_{\mathrm{BH}}>M^{\mathrm{crit}}_{\mathrm{BH}}$, so BH start evaporating. Due to evaporation, constraints are relaxed; hence, the collapse of the neutron star may not constrain quite massive DM particles, i.e., $m_{\chi}>6\times 10^{8}$ GeV for bosonic DM. 
Therefore, most of the DM parameter space constraints are relaxed from the evaporation constraint. The BH formation from BDM  constrains very limited parameter spaces $ 10^{-45} ~\mathrm{cm^{2}} \leq \sigma_{\chi \mathrm{N}} \leq 4\times 10^{-46} ~\mathrm{cm^{2}}$ of BDM masses $ 5\times 10^{7} ~\mathrm{GeV} \leq m_{\chi} \leq 5\times 10^{8}$ GeV. Further, we find that for FDM particles, BH formation happens on high DM masses, i.e. $m_{\chi}>10^{9}$ GeV. This is expected because overcoming the degeneracy pressure of the FDM requires more mass. Furthermore, BH formation in FDM does not constrain the DM microphysics as all the parameter space is relaxed from the saturation limit and evaporation constraints.

\begin{figure*}
  \includegraphics[height=3in,width=3.2in]{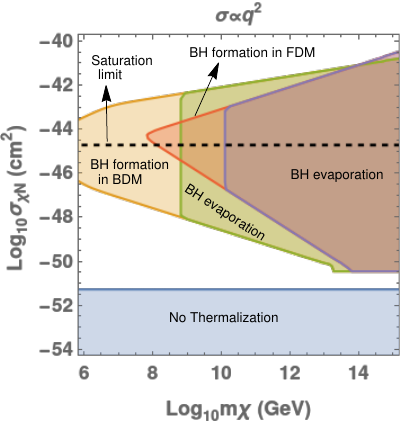}\hfil
 \includegraphics[height=3in,width=3.2in]{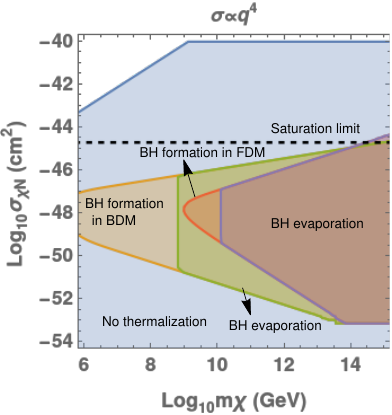}\par
\caption{Constraints on the DM-Nucleon scattering cross-section for momentum dependent cross-section ($\sigma\propto q^{2}$ (Left panel) and  $\sigma\propto q^{4}$ (Right panel) )
using the pulsar data J2124-3858. The representative colors carry the same meaning as earlier.}
\label{fig:mdconst}%
\end{figure*}

 Fig.~\ref{fig:mdconst} shows the constraint on the DM-nucleon scattering cross-section for the MDCS, case from the pulsar observation. The Left and Right panel corresponds to $\sigma\propto q^{2}$ ($N_{\mathrm{max}}=10$) and  $\sigma\propto q^{4}$ ($N_{\mathrm{max}}=1000$) cases, respectively. In this case, the DM accretion is enhanced (in comparison to the constant cross-section); therefore, the pulsar data constrain the massive DM particles. For $\sigma\propto q^{2}$, the pulsar data constraint  on cross-section $ 10^{-48} ~\mathrm{cm^{2}} \leq \sigma_{\chi \mathrm{N}} \leq 10^{-45} ~\mathrm{cm^{2}}$ for BDM masses $ 10^{6} ~\mathrm{GeV} \leq m_{\chi} \leq 6\times 10^{8}$ GeV and 
 $ 2\times 10^{-47} ~\mathrm{cm^{2}} \leq \sigma_{\chi \mathrm{N}} \leq 10^{-45} ~\mathrm{cm^{2}}$ for FDM masses $ 10^{8} ~\mathrm{GeV} \leq m_{\chi} \leq 10^{10}$ GeV. 
 Although the pulsar data also constrains more massive DM particles, those parameter spaces are relaxed by Hawking evaporation. Furthermore, for $\sigma\propto q^{4}$ case, DM accretion is even higher than  $\sigma\propto q^{2}$ case.
 However, in this model, the thermalization time is higher than the lifetime of the pulsar, which makes the BH formation highly unlikely, and therefore, this model does not constrain the DM parameters.  
\begin{figure*}
 \includegraphics[height=3in,width=3.2in]{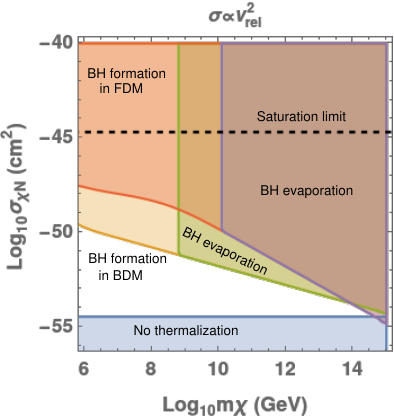}\hfil
 \includegraphics[height=3in,width=3.2in]{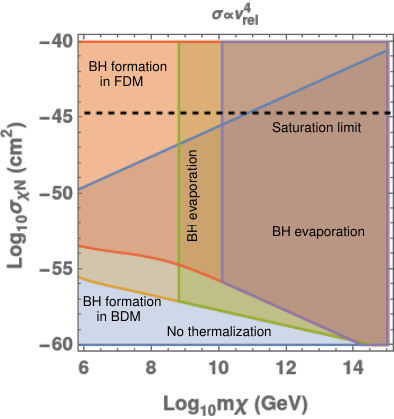}\par
 \caption{Constraints on the DM-Nucleon scattering cross-section for velocity-dependent cross-section ($\sigma\propto v^{2}_{\mathrm{rel}}$ (Left panel) and  $\sigma\propto v^{4}_{\mathrm{rel}}$ (Right panel) )
using the pulsar data J2124-3858. The representative colors carry the same meaning as earlier.}
\label{fig:vdconst}%
\end{figure*}
Furthermore, we constrain the velocity dependent DM-Nucleon scattering cross-section from the pulsar observation in Fig.~\ref{fig:vdconst}. The Left and Right panel corresponds to $\sigma\propto v^{2}_{\mathrm{rel}}$ ($N_{\mathrm{max}}=1000$) and  $\sigma\propto v^{4}_{\mathrm{rel}}$ ($N_{\mathrm{max}}=5000$) cases, respectively.  For $\sigma\propto v^{2}_{\mathrm{rel}}$, the captured DM particles are higher than the constant case; therefore, it constrains the DM parameter space more strongly. This model constrain $ 10^{-52} ~\mathrm{cm^{2}} \leq \sigma_{\chi \mathrm{N}} \leq 10^{-50} ~\mathrm{cm^{2}}$ for BDM masses, $ 10^{6} ~\mathrm{GeV} \leq m_{\chi} \leq 6\times 10^{8}$ GeV, and $ 10^{-50} ~\mathrm{cm^{2}} \leq \sigma_{\chi \mathrm{N}} \leq 2\times 10^{-47} ~\mathrm{cm^{2}}$ for FDM masses, $ 10^{6} ~\mathrm{GeV} \leq m_{\chi} \leq  10^{10}$ GeV. Furthermore, due to large positive power, the number of accreted DM particles is highest for model $\sigma\propto v^{4}_{\mathrm{rel}}$ (among all models discussed previously).  Therefore we expect that this model will put the tightest constraint. However, as seen in Fig.~\ref{fig:vdconst}, the thermalization constraint strongly restricts the DM parameter space and limits the cross-section, $ 10^{-50} ~\mathrm{cm^{2}} \leq \sigma_{\chi \mathrm{N}} \leq 10^{-47} ~\mathrm{cm^{2}}$ for the BDM masses, $ 10^{6} ~\mathrm{GeV} \leq m_{\chi} \leq 6\times 10^{8}$ GeV, and $  10^{-50} ~\mathrm{cm^{2}} \leq \sigma_{\chi \mathrm{N}} \leq 3\times 10^{-46} ~\mathrm{cm^{2}}$ for the FDM masses, $ 10^{6} ~\mathrm{GeV} \leq m_{\chi} \leq  10^{10}$ GeV. 

From the above discussion, we find that the pulsar data provide the strongest constraint on DM parameter space (both for bosonic and fermionic DM case) for the model $\sigma\propto v^{2}_{\mathrm{rel}}$.

\section{Conclusion}
\label{Sec:conc}
 A neutron star situated in a DM-dominated environment can efficiently capture the dark matter particles, and for large capturing, dark matter forms a black hole and destroys the host neutron star. In this work, we constrain the massive dark matter particles from the survival of the neutron star.

 To estimate the DM capture rate of massive dark matter particles $m_{\chi}>10^{6}$ GeV, we adapted the multiscatter capture formalism to a non-constant DM-nucleon scattering cross-section. Adapting a model-independent approach, we assumed a power law parametrization of the velocity dependent and momentum dependent DM-Nucleon scattering cross-section, $\sigma^{\mathrm{vd}}_{\chi\mathrm{N}}(v_{\mathrm{rel}})=\sigma_{\chi \mathrm{N}}\left(v_{\mathrm{rel}}/u_{0}\right)^{2\alpha}$ and $\sigma^{\mathrm{md}}_{\chi\mathrm{N}}(q)=\sigma_{\chi \mathrm{N}}\left(q/q_{0}\right)^{2\beta}$.
  Further, using the form factor and incorporating the general relativistic correction, we obtain analytic expressions for the capture rate in the neutron frame.
 Our estimation shows that the capture rate enhanced for the positive velocity and momentum dependent cross-section, and a number of the DM-Neucleon scattering, which leads to a black hole formation in a typical lifetime of the neutron star.    

 Further, applying the observation of pulsar data J2124-3858, we explored the constraint on the dark matter parameter space, i.e.,  $\sigma_{\chi \mathrm{N}} -m_{\chi} $ for non-constant cross-section. We obtained that pulsar data provide the strongest constraint on DM parameter space (both for bosonic and fermionic DM case) for the model $\sigma\propto v^{2}_{\mathrm{rel}}$. 
This model constrain $ 10^{-52} ~\mathrm{cm^{2}} \leq \sigma_{\chi \mathrm{N}} \leq 10^{-50} ~\mathrm{cm^{2}}$ for bosonic DM masses, $ 10^{6} ~\mathrm{GeV} \leq m_{\chi} \leq 6\times 10^{8}$ GeV, and $ 10^{-50} ~\mathrm{cm^{2}} \leq \sigma_{\chi \mathrm{N}} \leq 2\times 10^{-47} ~\mathrm{cm^{2}}$ for fermionic DM masses, $ 10^{6} ~\mathrm{GeV} \leq m_{\chi} \leq  10^{10}$ GeV. Interestingly, we also found that although for model $\sigma\propto v^{4}_{\mathrm{rel}}$, the DM accretion is higher, but due to a large thermalization time, it put weak constraints on the DM parameters. Furthermore, a large momentum dependent cross-section model, i.e., $\sigma\propto q^{4}$ provides no constraint on the DM parameters. 

To the best of our knowledge, this is the first constraint on the massive DM particles from the neutron star survival using the velocity and momentum dependent cross-section (in the multiscatter capture formalism). Nevertheless, we also point out that our calculation of capture rate has been done under certain assumptions, such as non-relativistic scattering of DM-nucleon, saturation cross-section is constant with DM mass (however, varies for massive DM particle), nucleon density  (also escape velocity) is constant. After inclusion, these effects will change the constraint, and we leave these interesting aspects for the future.

\section{ACKNOWLEDGEMENT}
This work is supported by the National Natural Science Foundation of China (NNSFC) under grants No. 12275134, No. 12335005, and No. 12147228. AKM would like to thank Prof. Lei Wu and Prof. Chih-Ting Lu for the useful discussions. 

\appendix
\section{Calculation of multiscatter capture of DM particles via constant cross-section}
\label{constant}
After having equipped with the probability expression (see Section), we now estimate the multiscatter capture rate using Eq.(\ref{eq:Cnx}). The derivation of Eq.(\ref{eq:Cnx}) has been done is the assumption of multi-capture with $v_{esc} (r)\simeq v_{esc} (R_{\mathrm{NS}}) $ and number density of the baryons are constant inside the neutron star, $n_{B}(r)=n_{B}$ see Ref. \cite{Bramante:2017xlb}. To estimate the capture rate, we apply the Maxwell distribution function in the neutron star frame, Eq. (\ref{eq:dist}).

In a constant cross-section case, the probability, $P^{\mathrm{cont}}_{\mathrm{N}}(\tau_{0})$ is independent of the DM velocity. So, applying Eq.(\ref{eq:Cnx}), Eq.(\ref{eq:dist}), and Eq.(\ref{eq:pconst}) the analytic expression of $C_{\mathrm{N}}$ is obtained as
 \begin{equation}
  C^{\mathrm{cont}}_{\mathrm{N}}(\tau)=   \frac{\pi R^{2}}{1-v^{2}_{\mathrm{esc}}}\frac{1}{6\pi u_{0}}\frac{\rho_{\chi}}{m_{\chi}}P^{\mathrm{cont}}_{\mathrm{N}}(\tau_{0})f(v_{\mathrm{N}},\eta)\Theta(v_{\mathrm{N}}-v_{\mathrm{esc}})
   \label{eq:cnconst}
 \end{equation}
 where 
\begin{align}
& f(v_{\mathrm{N}},\eta)= \frac{1}{\eta}\left[e^{-\frac{3(v^{2}_{\mathrm{N}}-v^{2}_{\mathrm{esc}})}{2u^{2}_{0}}}+\frac{2\sqrt{v^{2}_{\mathrm{N}}-v^{2}_{\mathrm{esc}}}~\eta}{u_{0}} +\eta^{2}\right]
\times \bigg[-\sqrt{6\pi}u_{0}\left( -1+\frac{2\sqrt{v^{2}_{\mathrm{N}}-v^{2}_{\mathrm{esc}}}\eta}{u_{0}}\right)\sqrt{v^{2}_{\mathrm{N}}-v^{2}_{\mathrm{esc}}}\nonumber \\
&
\left(1+e^{\frac{\sqrt{(v^{2}_{\mathrm{N}}-v^{2}_{\mathrm{esc}})}\eta}{u_{0}}}-2e^{-\frac{3}{{2u^{2}_{0}}}\left[(v^{2}_{\mathrm{N}}-v^{2}_{\mathrm{esc}})+2\sqrt{(v^{2}_{\mathrm{N}}-v^{2}_{\mathrm{esc}}}u_{0}\eta\right]}\right)u_{0}\eta 
+\pi\left(3v^{2}_{\mathrm{esc}} +u^{2}_{0}(1+3\eta^{2})\right)e^{\frac{3}{{2u^{2}_{0}}}\left[v^{2}_{\mathrm{N}}-v^{2}_{\mathrm{esc}}+2\sqrt{v^{2}_{\mathrm{N}}-v^{2}_{\mathrm{esc}}}u_{0}\eta+u^{2}_{0}\eta^{2}\right]}
\nonumber \\
&
 \left\{ ( 2 \mathrm{Erf} \sqrt{\frac{3}{2}\eta} + \mathrm{Erf} \left(\frac{\sqrt{\frac{3}{2}}(\sqrt{v^{2}_{\mathrm{N}}-v^{2}_{\mathrm{esc}}}-u_{0}\eta)}{u_{0}}\right) 
 + \mathrm{Erf} \left(\frac{\sqrt{\frac{3}{2}}(\sqrt{v^{2}_{\mathrm{N}}-v^{2}_{\mathrm{esc}}}+u_{0}\eta)}{u_{0}}\right)\right\} \bigg] , 
 \label{eq:f0}
 \end{align}
 where, $v_{\mathrm{N}}=v_{\mathrm{esc}}\left(1-\frac{\beta_{+}}{2} \right)^{-\mathrm{N}/2}$ is the DM velocity after $N$ scattering, and $\mathrm{erf(z)}=\frac{2}{\sqrt{2}} \int^{z}_{0}e^{-y^{2}}dy$ represent the error function.
In the above equation, $\rho_{\chi}$ is the DM energy density surrounding the NS, and $N_{\mathrm{B}}$ is the total number of baryons inside the NS. We checked that in the rest frame of DM halo, i.e., $\eta \rightarrow 0$, and point-like structure of nucleon (neglecting the form factor), our expression of $C^{\mathrm{const}}_{\chi\mathrm{N}}$ matches with the Ref.~\cite{1987ApJ...321..571G}.

\section{Derivation of multiscatter capture of DM particles via MDCS}
\label{momdep}
In this case, $\sigma^{\mathrm{md}}_{\chi\mathrm{N}}\propto q^{2\beta}$, where $q\simeq \sqrt{2}\left(\frac{m_{\chi}m_{\mathrm{N}}}{m_{\chi}+m_{\mathrm{N}}}\right)v_{\mathrm{esc}}$.  So, $P^{\mathrm{MDCS}}_{\mathrm{N}}(\tau_{0},q)$ is independent of the initial DM velocity. Using this, we obtain the analytic form of the capture rate after $N$ scattering for the arbitrary momentum dependence cross section. The form of $C_{\mathrm{N}}$ in the MDCS case is given by
\begin{equation}
  C^{\mathrm{MDCS}}_{\mathrm{N}}(\tau_{0},q)=   \frac{\pi R^{2}}{1-v^{2}_{\mathrm{esc}}}\frac{1}{6\pi u_{0}}\frac{\rho_{\chi}}{m_{\chi}}P^{\mathrm{MDCS}}_{\mathrm{N}}(\tau_{0},q)f(v_{\mathrm{N}},\eta)\Theta(v_{\mathrm{N}}-v_{\mathrm{esc}})
\label{eq:cnmdcs}
 \end{equation}

\section{Derivation of multiscatter capture of DM particles via VDCS}
\label{veldep}
In this case, $\sigma^{\mathrm{md}}_{\chi\mathrm{N}}\propto w^{2\beta}$, where $w=\sqrt{u^{2}+v^{2}_{\mathrm{esc}}}$. Therefore the probability $P^{\mathrm{VDCS}}_{\mathrm{N}}(\tau_{0},w)$ depends on the DM velocity, see Eq. (\ref{eq:pvdcs}). Using this probability expression, no reliable analytic (or numerical) solution was obtained. Therefore, to calculate the capture rate, we follow a conservative approach and approximate  $w\simeq w_{0}=\sqrt{u^{2}_{0}+v^{2}_{\mathrm{esc}}}$ in the probability expression. After applying this approximation, we can easily integrate and obtain an analytic expression for the capture rate as
\begin{equation}
 C^{\mathrm{VDCS}}_{\mathrm{N}}(\tau_{0},w_{0})=   \frac{\pi R^{2}}{1-v^{2}_{\mathrm{esc}}}\frac{1}{6\pi u_{0}}\frac{\rho_{\chi}}{m_{\chi}}P^{\mathrm{VDCS}}_{\mathrm{N}}(\tau_{0},w_{0})f(v_{\mathrm{N}},\eta)\Theta(v_{\mathrm{N}}-v_{\mathrm{esc}})
\label{eq:cnvdcs}
 \end{equation}

\bibliographystyle{utphys}
\bibliography{BDMinNS}
 
\end{document}